# Crystalline Formations of NbN/4H-SiC Heterostructure Interfaces

*Michael B. Katz[†], Chieh-I Liu[†,§], Albert F. Rigosi[†], Mattias Kruskopf[†,‡], Angela Hight Walker[†], Randolph E. Elmquist[†], and Albert V. Davydov[†*]*

[†]National Institute of Standards and Technology, Gaithersburg, MD 20899, United States

[§]Department of Chemistry and Biochemistry, University of Maryland, College Park, MD 20742, United States

[‡]Physikalisch-Technische Bundesanstalt, Bundesallee 100, 38116 Braunschweig, Germany

ABSTRACT Given the importance of incorporating various superconducting materials to device fabrication or substrate development, studying the interface for possible interactions is warranted. In this work, NbN films sputter-deposited on 4H-SiC were heat-treated at 1400 °C and 1870 °C and were examined with transmission electron microscopy to assess whether the interfacial interactions undergo temperature-dependent behavior. We report the diffusion of NbN into the SiC substrate and the formation of NbN nanocrystallites therein during the 1400 °C treatment. After the 1870 °C treatment, tiered porosity and the formation of voids are observed, likely due to catalytic reactions between the two materials and accelerated by the stresses induced by the differences in the materials' coefficients of thermal expansion. Lastly, Raman spectroscopy is employed to gain an understanding of the interface lattices' optical responses.

INTRODUCTION

Since its discovery, superconducting NbN has played a key role in several facets of materials science and device fabrication.[1] The modern use of NbN is sustained by the ease with which it can be deposited on a suitable substrate as well as its ability to maintain a superconducting state until above 10 K and for some considerable amount of magnetic flux density (approximately 10 T).[2] Thin films of NbN exhibit some versatility when applying them to quantum computing in the form of single photon detectors, flux quantum circuits, and qubits.[3,4,5] Furthermore, these films and their similar counterparts can be deposited onto a variety of substrates including SiC for the purposes of vertical transistors,[6,7,8,9,10] resistance standards,[11,12] and Josephson junction devices.[13]

Given the extent of functionality for these thin films, knowledge on the interactions between NbN and its corresponding substrate is of vital importance for device engineering. Among available substrates used for material studies, SiC stands out in the following ways: wide bandgaps for semiconductor applications, low loss for THz frequencies, chemical compatibility with other thin films (from insulating to superconducting) fit for quantum communications, and wide commercial availability.[14] For these reasons, interest remains strong in understanding the interfacial interactions between NbN and SiC. Results for NbN/3C-SiC interactions have been recently reported assessing the viability of this heterostructure for hot-electron bolometer mixers and other THz applications.[15,16] Similarly, results as recent as last year reported on the interactions between NbN and 6H-SiC.[10,17] Given this new interest in examining SiC polymorphs and their interactions with thin NbN, it is only fitting that work should be done using 4H-SiC. 4H-SiC is another polymorph known to have superior electronic properties, such as a higher electron and hole mobility as well as a higher Baliga figure of merit,[18] when compared to polymorphs 3C-SiC and 6H-SiC.[19,20]

In this work, NbN films were deposited on 4H-SiC and heat treated at 1400 °C and 1870 °C. The heat-treated and as-deposited films were examined with transmission electron microscopy (TEM), X-ray diffraction (XRD), scanning electron microscopy (SEM), and Raman spectroscopy to observe the temperature dependence of any interfacial reactions. All depositions were performed via sputtering as with similar processes,[21,22,23] but it should be noted that there are several varieties of methods for depositing NbN.[2,24,25,26,27] The diffusion of Nb and N into the 4H-SiC substrate and subsequent recrystallization therein is reported at the lower of the two heat treatment temperatures. For the higher temperature, we observe tiered porosity and the formation of voids, likely resulting from the differences between the coefficients of thermal expansion (CTEs) of the interface materials. Data acquired with Raman spectroscopy suggest the formation of carbon-based lattices at the interface for both treatment temperatures.

RESULTS AND DISCUSSION

*Annealing Summary and As-Deposited Samples*

Various 4H-SiC substrates were diced from a wafer and subsequently placed in a sputter chamber. After the NbN layer was deposited, specimens were either left as grown, heated to 1400 °C, or heated to 1870 °C. Each of the three specimens were then prepared by focused ion beam (FIB) for examination by transmission electron microscopy (TEM), as described in detail in the Methods section. Figure 1 summarizes the results schematically as they will be shown in the following sections. In the case of a lower-temperature 1400 °C film, diffusion of NbN and the formation of NbN crystallites and voids within the near-interface region were observed, whereas in the higher-temperature 1870 °C film, larger voids with less surface area formed. In both films, the NbN formed into grains generally tens of microns in extent.

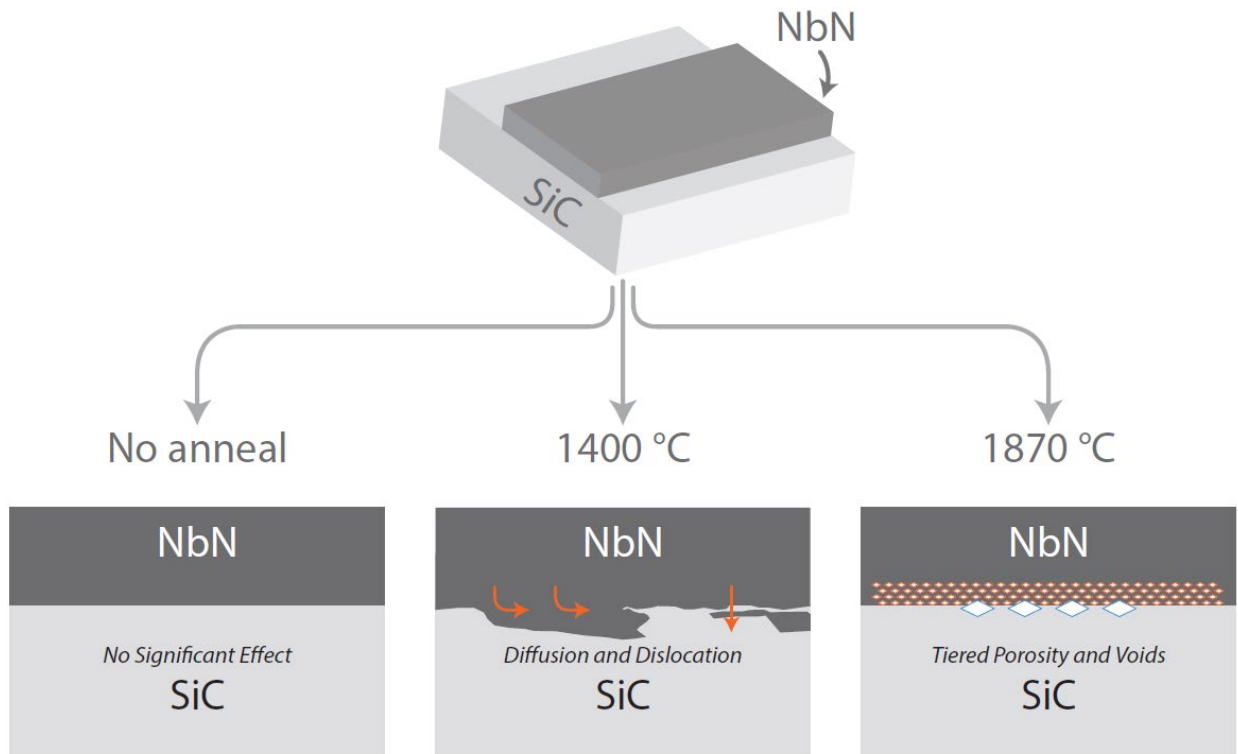

**Figure 1.** Schematic illustration of the NbN–4H-SiC interface after deposition and the two heat treatments. For (a) the as-deposited interface, no significant reactions were observed. At 1400 °C, some of the NbN appears to diffuse into the SiC substrate from the film, and forms crystallites. At 1870 °C, the NbN film coarsens into larger grains, resulting in a porous film-substrate interface.

In the first case, the as-deposited sample was examined for a baseline comparison to annealed samples. XRD data were acquired and shown in Fig. 2 (a), with the data suggesting that the NbN film is amorphous. Fig. 2 (b), shows a TEM image of the overall polycrystalline NbN film atop the 4H-SiC substrate, with the inset SAED pattern for the NbN film corroborating the polycrystallinity of the NbN.

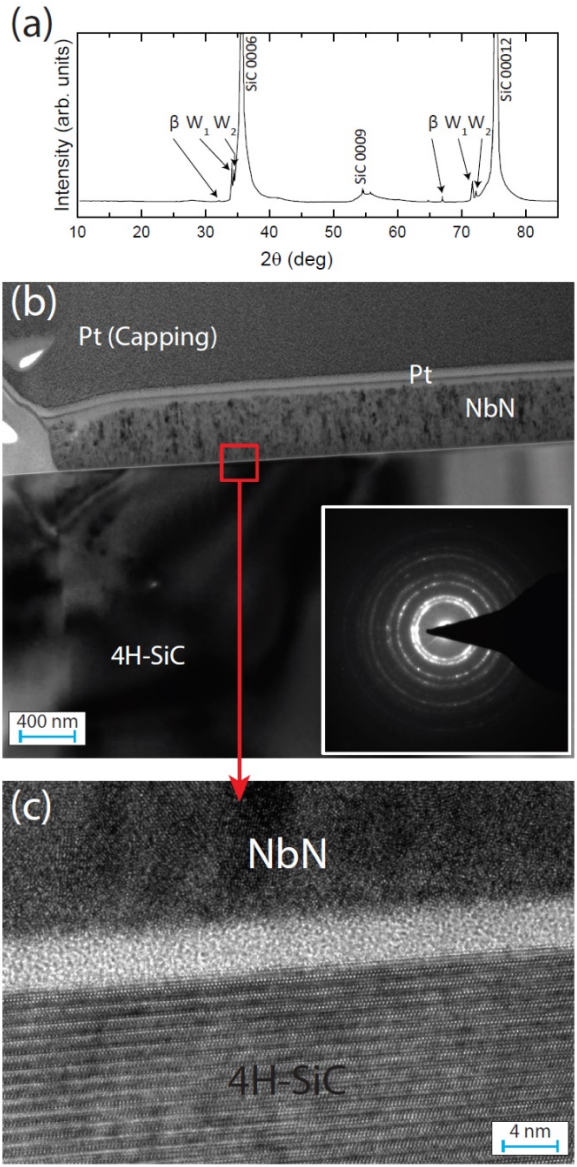

**Figure 2.** (a) XRD data taken on NbN suggesting an amorphous structure. (b) A TEM image cross-section of the NbN/4H-SiC film stack, with a polycrystalline SAED pattern for the NbN film inset. (c) A higher-magnification TEM image of the indicated region in (b) showing the interface more clearly

*Heat Treatment at 1400 °C*

Heating the film to 1400 °C induces a trio of major reactions, both in the NbN film and the 4H-SiC substrate. The extent of the transformations is shown in Figure 3. The NbN film has completely recrystallized into large grains from its formerly nano-crystalline state. The grains are approximately 100 nm to 300 nm in extent, with single grains spanning the full depth of the film. Also immediately evident, in contrast to the as-grown film, is the separation between the NbN film and the 4H-SiC substrate. It appears not simply delaminated but very rough and uneven at the length scale of the film. This is notable because SiC is generally stable under heat treatments at much higher temperatures. This decomposition of the interfacial region of the 4H-SiC substrate may be partially accounted for by the tendency to relieve increased interfacial energy due to a large mismatch in the CTE during cooling.[28][29][30]

The third reaction, though, may offer an additional explanation for the extensive degradation of the 4H-SiC at the interface. Examining the near-interfacial region of the 4H-SiC at high magnification reveals a population of faceted nanocrystallites distinct in contrast from the surrounding SiC material. EDS reveals these crystallites to comprise niobium and nitrogen either in part or in full and is likely some phase of $Nb_xN_y$. The presence of these crystallites and constituent species within the transformed regions of the 4H-SiC suggest the possibility that the decomposition of the erstwhile 4H-SiC surface was catalyzed by the presence of and facilitated the migration of these crystallites away from the film and into the substrate.

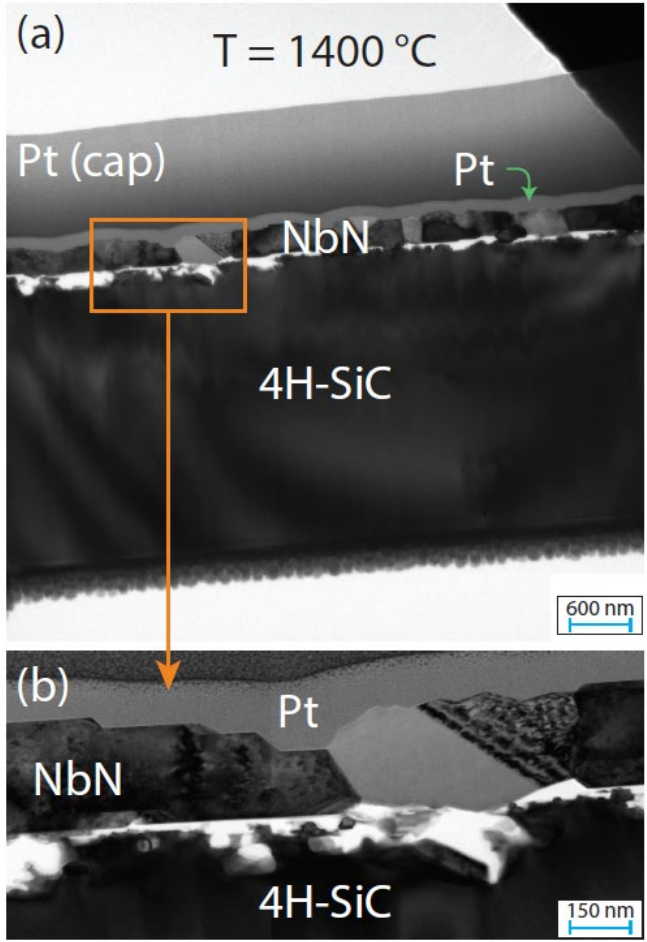

**Figure 3.** (a) Overview of a cross-section of the film heat-treated at 1400 °C by TEM and (b) a higher-magnification image of the area called out in (a). The coalescence and subsequent faceting of the NbN film is especially apparent in the grain to the left side of the image. The regions of bright contrast between the NbN film and the 4H-SiC substrate are porosity induced by the heat treatment.

In the XRD data shown in Fig. 4 (a), three distinct crystalline NbN phases are present: (I) δ-NbN (225), which has a characteristic lattice constant $a = 0.446$ nm (II) tetragonal $Nb_4N_3$ (139), with $a = 0.438$ nm and $c = 0.863$ nm, and (III) a primitive cubic $NbN_x$, with $a = 0.694$ nm, which has scant reports in the literature.[31] Fig. 4 (b), shows the Nb-N crystallites in detail.

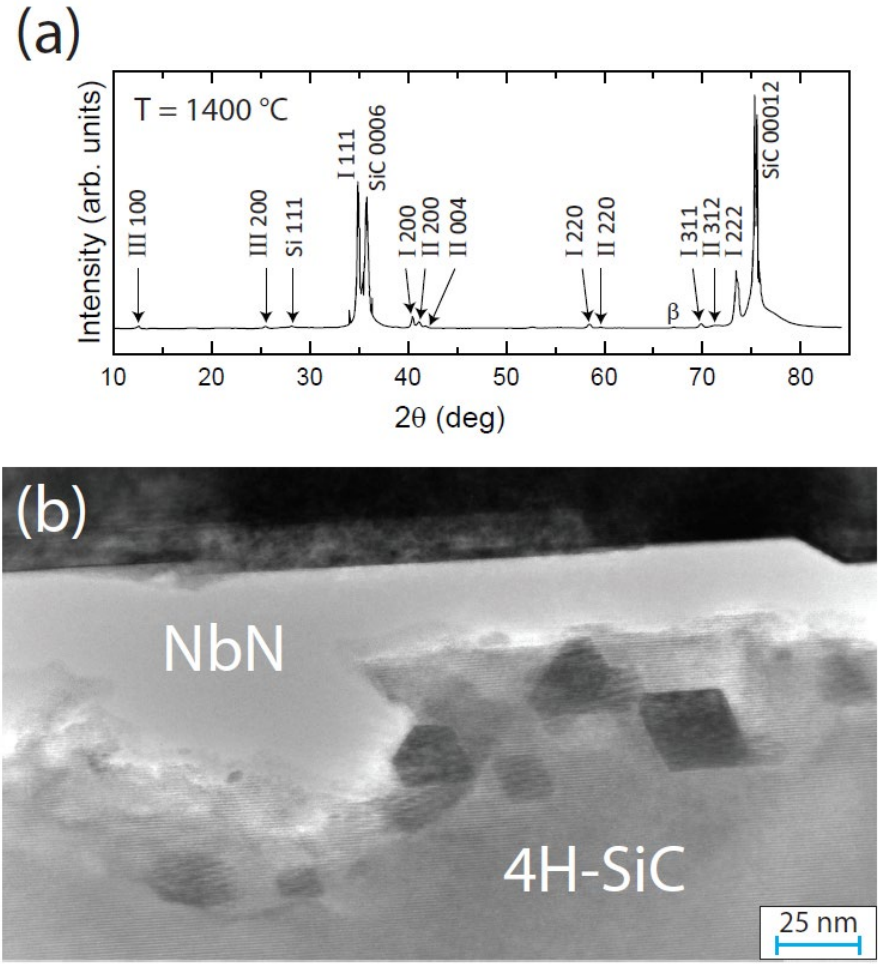

**Figure 4.** (a) XRD data show a mixture of three crystalline NbN phases: (I) δ-NbN (225), (II) tetragonal $Nb_4N_3$, and (III) cubic primitive $NbN_x$. The cubic response is very small compared with the other two phases. (b) A high-resolution TEM image of the near-surface region of the 4H-SiC substrate revealing NbN crystallites having formed after migration of Nb and N from the film but before the voids opened up.

*Heat Treatment at 1870 °C*

Increasing the heat treatment temperature to 1870 °C results in a more well-defined and locally smooth interface, but still features significant intrusion into the 4H-SiC surface by a newly recrystallized NbN film. Cross-sections of this specimen are shown in Figure 5 (a and b) In a similar fashion to the 1400 °C case, the NbN film has recrystallized into grains of about 300 nm to 500 nm in extent and through the full thickness of the film. Closer inspection reveals the well-faceted nature of the recrystallized NbN. The insets in Figure 5 (b) show clear preferences in the NbN for recrystallizing along low-index planes, both at the NbN/4H-SiC interface and at the NbN/air interface. The scale of the faceting can be seen more clearly in Figures 5 (c and d), in which the film coarsened into roughly equiaxed grains in which the grains formed terraces to favor the (111) plane interface with air. Also evident are large pores, of order 100 nm, between some coarsened grains which are too dispersed to have been captured in the relatively small area of the TEM specimen.

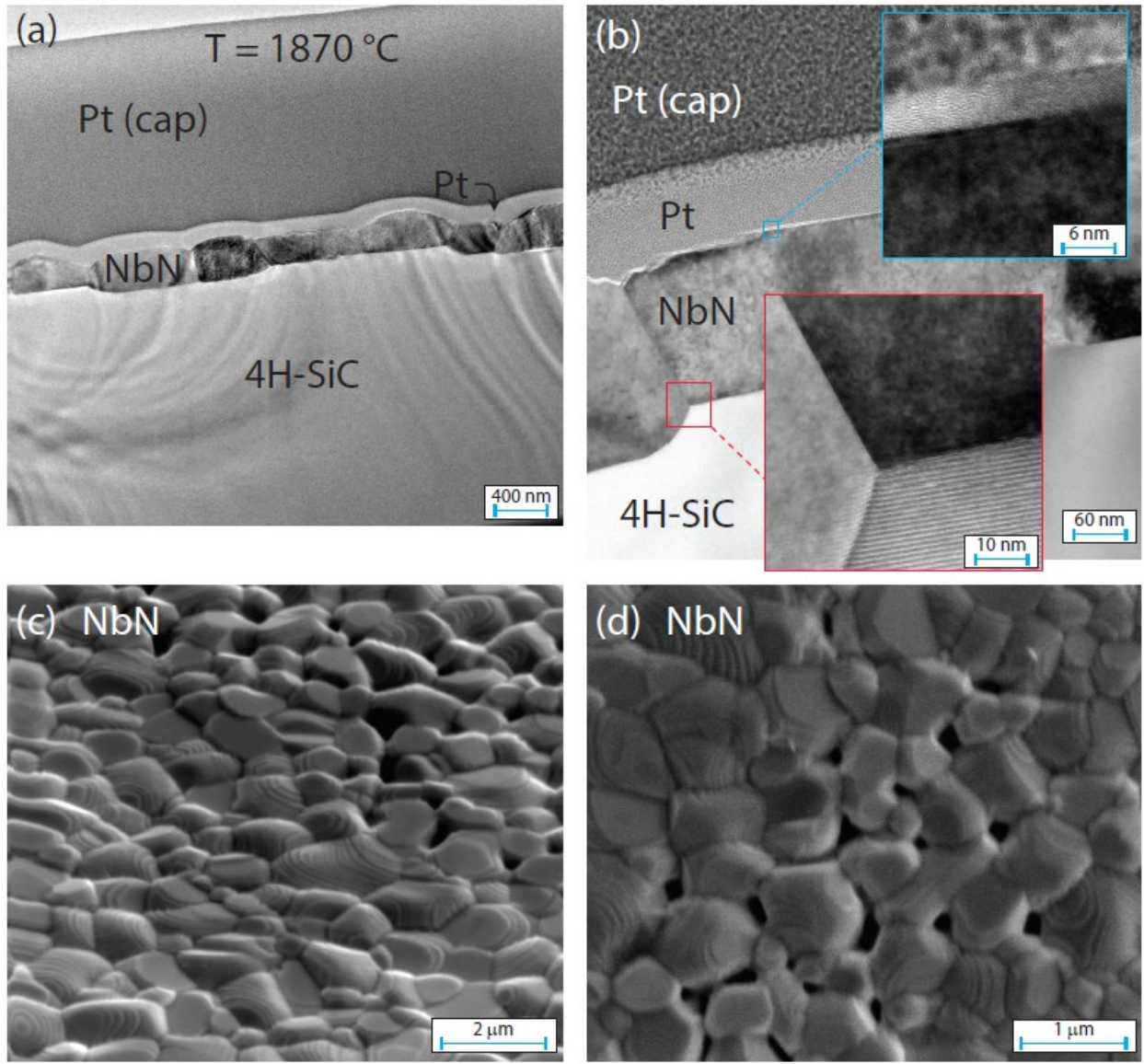

**Figure 5.** (a) TEM image of the cross-section of the NbN/4H-SiC interface. (b) A magnified region of the interface shows how well-faceted the intersection is between NbN domains and the 4H-SiC substrate. A secondary magnification is shown in red for the intersection and blue for the interface between NbN and the Pt capping layer. SEM images of the NbN film surface at (c) oblique and (d) surface-normal angles reveal the NbN grains faceting into step-like terraces. Large voids can be seen between some NbN grains, especially in (d).

To provide more evidence of NbN becoming crystalline, the regions displayed in Fig. 6 (a) were examined more closely. Vacuum voids formed regularly along the NbN/4H-SiC interface, which is a response less likely to occur when one of the two interacting materials is amorphous. For instance, the thin Pt layer, owing to its more amorphous structure, fills the contour of its neighboring layer, whereas in the case of the NbN/4H-SiC, having two crystalline structures competing for long-range order yields vacuum regions that may be sustained by the mechanical strength of the adjoining layers. This explanation is further justified by the differing CTCs of the two interface materials.[26-28] In Fig. 6 (b) and (c), SAED images corresponding to the 4H-SiC and NbN are shown to provide evidence of the crystallinity of both materials, respectively. In the case of the NbN film, this image gives supporting evidence of the film's crystalline nature post-anneal and may be compared with the one shown in the inset of Fig. 2 (a), where the NbN film does not depart from its amorphous composition. Fig. 6 (d) shows the XRD data, and unlike the mixed phases seen in the 1400 °C anneal case, this pattern, as well as the SAED patter, shows the cubic phase of NbN.

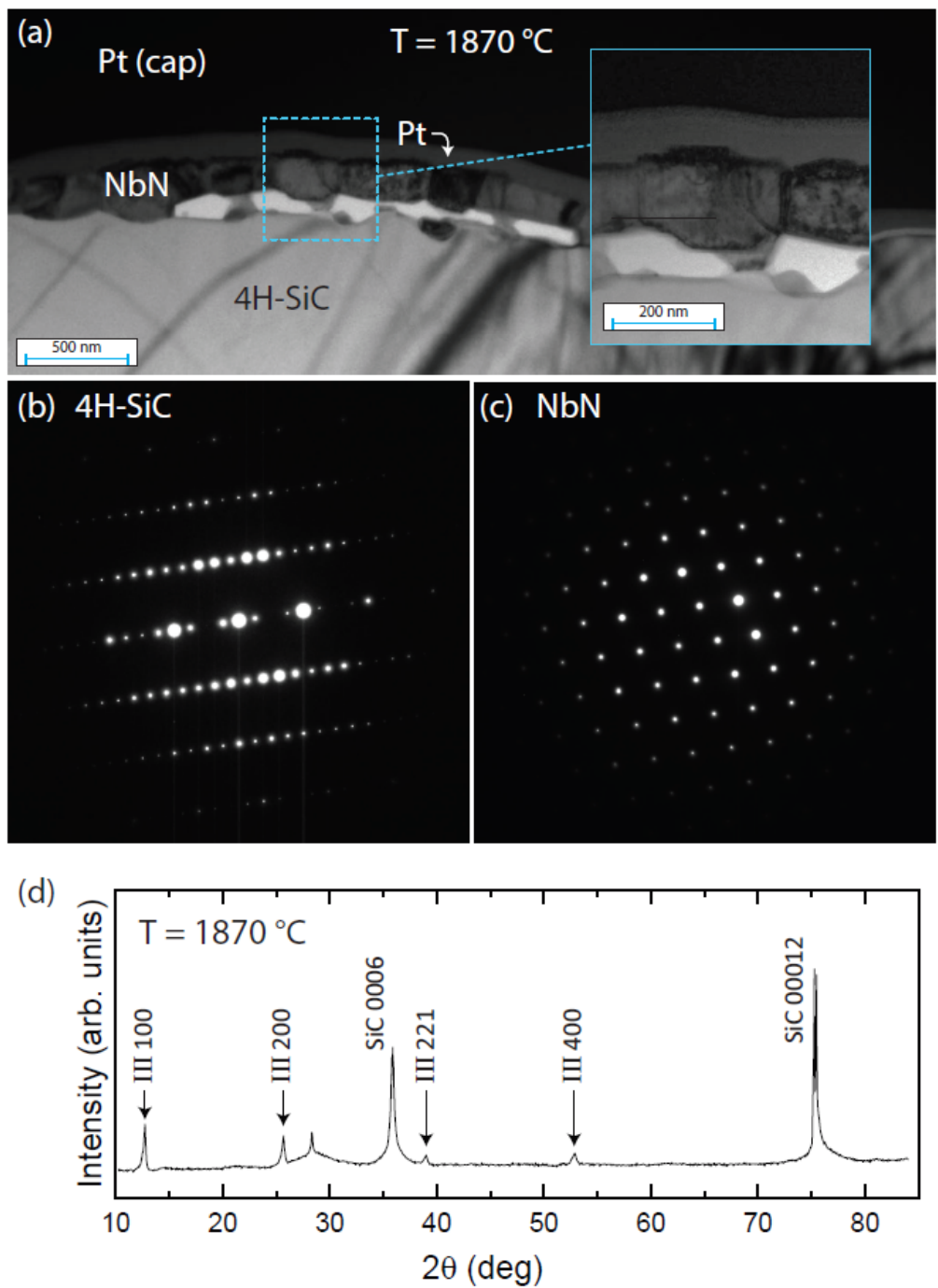

**Figure 6.** (a) Along portions of the NbN/4H-SiC interface, vacuum voids form from the differing CTCs. The magnified region in blue reveals the well-defined crystalline structure taken on by the NbN film. (b) SAED image of the 4H-SiC is shown to provide evidence of the substrate crystallinity. (c) SAED image of the NbN film is shown to give supporting evidence of the film's

crystalline nature post-anneal. This can be compared with Fig. 2 (a), where the NbN film appears amorphous. (d) XRD data showing a dominant formation of $NbN_x$.

*Section 4 – Raman and Carbon Growth*

Raman spectroscopy was performed on the samples to get a more comprehensive understanding of the interactions within the interface after the annealing process. Due to the opaque NbN film, the excitation laser was sent through the bottom of the substrate, as has been done in other work to reduce the contribution from the dominant SiC response.[32][33] Rectangular area Raman acquisition maps were also collected across the samples with step sizes of 20 μm in a 5 by 3 raster-style grid, as shown in Fig. 7 (a). For each spatial point, a second measurement was taken with a laser focus away from the interface and in the 4H-SiC in order to obtain a pure 4H-SiC signal, as illustrated in Fig. 7 (a). The resulting datasets yielded two types of responses, a response from the interface (which includes a response from the 4H-SiC) and one from the 4H-SiC bulk (red and black curves in Fig. 7 (b), respectively). A sloped background, resulting from the metallic NbN, was subtracted from all the data. The highest contributing Raman responses for NbN have been measured in other work.[27]

There is a quantifiable difference between the two types of responses, so the differences of the spectra were tabulated and shown in Fig. 7 (c) for 1400 °C and 1870 °C in the top and bottom panel, respectively. Within each panel, five example subtracted spectra are shown along with a solid averaged curve in light blue. The observed differences in the spectra can be attributed to the vibrational density of states (VDOS) of a proto-graphene layer known as the interfacial buffer layer (IBL), which is still covalently bonded to the 4H-SiC.[34][35] One of the only significant differences between the analyses for the two temperatures is the emergence of an additional contributing peak at 1870 °C near 1365 cm$^{-1}$, which may immediately come to mind as a D peak for graphene.

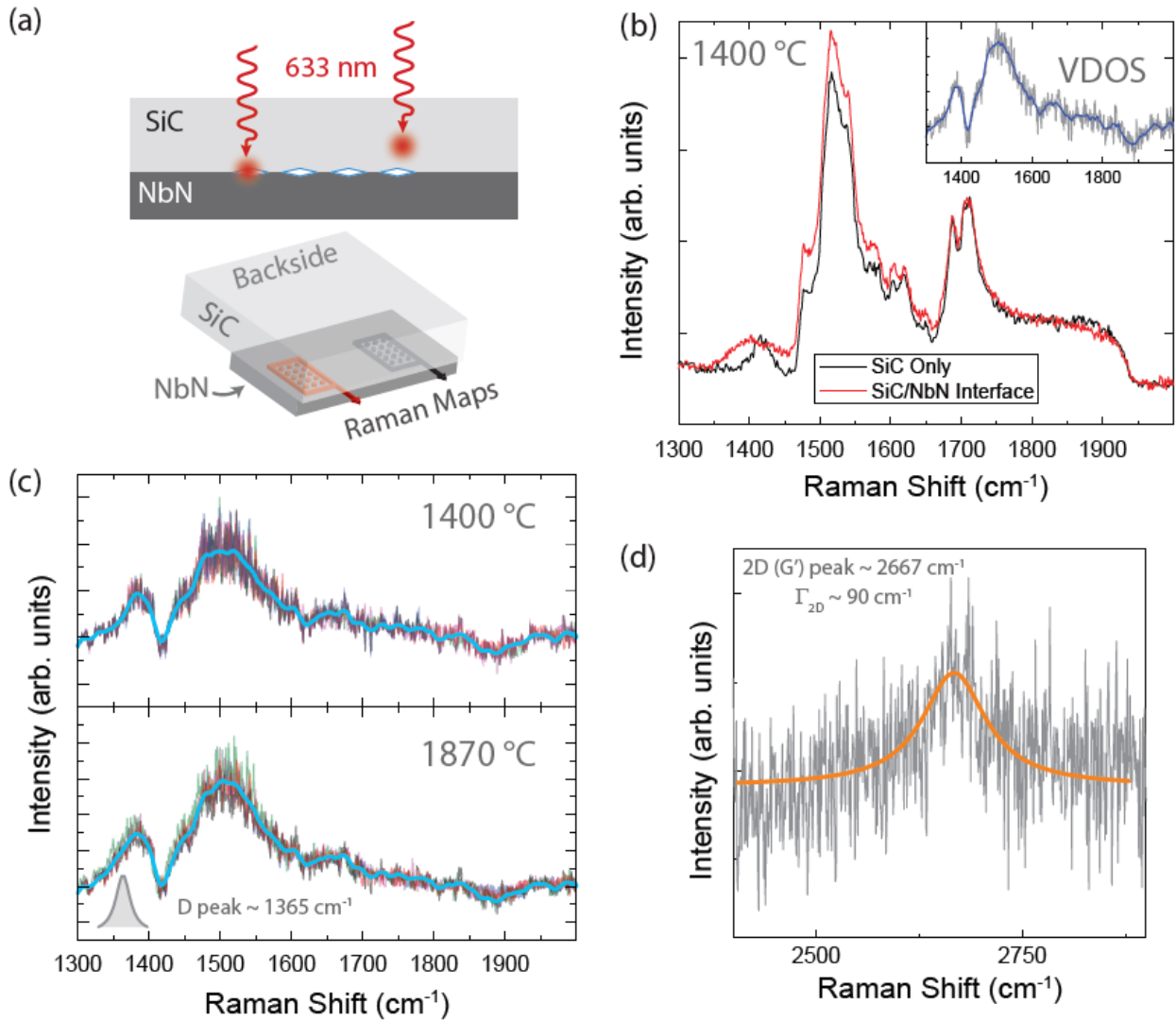

**Figure 7.** (a) Illustration of the Raman measurement setup, where an incoming laser is focused through the backside of the 4H-SiC substrate on both the interface and the 4H-SiC bulk. (b) An example pair of datasets from the corresponding Raman maps from the 1400 °C sample showing a measurable difference between the interface/4H-SiC and pure 4H-SiC (in red and black, respectively). The difference is measurable for all spectra and an example subtracted spectrum is shown in the inset. (c) The differences are attributable to the buffer layer seen when carbon-based lattices form on the surface of 4H-SiC. The top and bottom panels show the buffer layer VDOS for 1400 °C and 1870 °C, respectively. The bottom panel also contains evidence of a D peak at 1365 cm$^{-1}$. (d) In approximately 20 % of acquired spectra taken at 1870 °C, a weak 2D peak emerged, indicating the formation of a new lattice. A 20-pt adjacent averaged, smoothed curve is shown in orange.

The attribution is more nuanced since there are several crystal formations that can yield similar Raman data. For instance, both NbC and Si$_3$N$_4$ have D, G, and 2D peaks in similar locations to carbon-based lattices.[36,37,38] To provide additional evidence for why the response may be attributed as carbon-based, the spectral neighborhood of the 2D peak was analyzed for 1870 °C (all 1400 °C data in this spectral neighborhood showed no measurable responses). The major issue with this approach was that a majority of the acquired spectra contained no evidence of a 2D peak at all. A 2D-like response was only visible in about 20 % of the collected spectra, and an example of this visibility is shown in Fig. 7 (d). There are several facts to note. First, since the 2D peak was not visible for all spectra, a very high level of disorder of the new lattice should be expected. Second, this disorder was further validated by the relatively strong D peak (that is, the ratio of $I_D/I_{2D}$ is much greater than 1). Third, there was a complete lack of additional peaks typically seen in niobium-based or silicon-based lattices in the same neighborhood with similar intensities as the 2D peak. These observations, plus the agreeing data from the VDOS, indicate that a carbon-based lattice was the dominant new formation at the interface. These analyses provide additional insights into the interactions and changes that have occurred at the NbN/4H-SiC interface.

CONCLUSION

In this work, depositions of NbN films on 4H-SiC were examined with transmission electron microscopy (TEM) and other related techniques after heat treatments at 1400 °C and 1870 °C as well as as-grown to assess the extent of interfacial interactions and any possible temperature-dependent behavior. Overall, diffusion of NbN into and the nucleation of crystallites of NbN within the near-surface regions of the 4H-SiC substrate was observed at 1400 °C, whereas at 1800 °C, tiered porosity and the formation of voids were observed, likely resulting from the differences between the CTCs of the interface materials. XRD data revealed the high order of cubic $NbN_x$ that formed in the 1800 °C case. Additionally, Raman measurements showed the effects on the 4H-SiC by revealing a carbon-based lattice formation at the interface. These observations have clarified the extent to which interfacial interactions are temperature dependent.

METHODS

SiC substrates were prepared and diced from on-axis 4H-SiC(0001) semi-insulating wafers as in other similar work.[39]

A Denton Vacuum Discovery 550 [see Acknowledgments] was used to deposit NbN on 4H-SiC substrates. The process began when the minimum vacuum of $1.33 \times 10^{-3}$ Pa was met. Flow rates correspond to gases at standard temperature and pressure. The deposition process for superconducting NbN contains seven steps. (1) Nitrogen and argon gas flow at 120 $cm^{-3}$/min and 20 $cm^{-3}$/min, respectively, for 20 min (to clean the chamber). (2) Pre-sputtering ensures removal of possible gaseous contaminants on the sputter targets. For 10 min, argon gas flows at 50 $cm^{-3}$/min while the Nb sputter target is being cleaned (DC ion source, 2 W). (3) A second pre-sputtering step with nitrogen and argon gas flow at 3 $cm^{-3}$/min and 50 $cm^{-3}$/min, respectively, for 5 min (Nb DC ion source, 1.5 W). (4) Nb sputtered in the same conditions as step 3 to promote growth of superconducting NbN (10 min). (5) Chamber flush for 1 min using argon gas flow at 50 cm$^{-3}$/min. (6) Pre-sputter for 5 min using argon gas flow at 50 cm$^{-3}$/min (Pt DC ion source, 0.2 W). (7) Pt is deposited as a protective capping layer to protect it from oxidation (same conditions as step 6). This deposition yields: 330 nm NbN, and 50 nm Pt. In corresponding TEM images, a significantly thicker, second Pt layer is deposited as part of the TEM sample preparation method.

The annealing step at 1875 °C was performed with a background argon gas environment (originating from a 99.999 % liquid argon source) at 104 kPa (slightly above atmospheric pressure). The ramping rate was approximately 90 °C/min, and the anneal began with a gradual heating to 1050 °C followed by the second segment of heating to 1875 °C. This process occurs over about 10 minutes. The cooling process occurs after annealing at the maximum temperature for 4.5 min, with a cooling rate of 90 °C/min. The annealing is performed with a graphite-lined resistive-element furnace (Materials Research Furnaces LLC [see Acknowledgment]) that is cooled by chilled water. Pumping and backfilling is done with a mass-flow valve and a multi-stage roots dry pump. For samples annealed at 1400 °C, all the same conditions apply except for the maximum temperature and the hold time at maximum temperature, which in this case, was 30 min. Some samples were not annealed at all.

Raman measurements were performed with a Renishaw InVia micro-Raman spectrometer [see Acknowledgement] using a 633 nm wavelength excitation laser source. The spectra were collected using a backscattering configuration, 300 s acquisition time, 1 μm spot size, 50 × objective, 1.7 mW power, and a 1800 mm$^{-1}$ grating. Rectangular acquisition maps were also collected with step sizes of 20 μm in a 5 by 3 raster-style grid.

Cross-sectional TEM specimens were prepared by focused ion beam milling in a FEI Nova NanoLab DualBeam FIB using a gallium ion beam. Pt capping layers were used to protect the film from the ion beam during processing. TEM observation was performed in a FEI Titan TEM under a 300 kV accelerating voltage.

AUTHOR INFORMATION

**Corresponding Author**

* Correspondence to: albert.davydov@nist.gov

**Author Contributions**

C.-I.L, R.E.E, and M.K. prepared samples. M.B.K. performed electron microscopy measurements and analysis. A.V.D. performed x-ray diffraction and analysis. A.F.R., A. R. H. W., R.E.E., and A.V.D. assisted with the analyses, support, and general project oversight. The manuscript was written through contributions of all authors. All authors have given approval to the final version of the manuscript.

**Funding Sources**

Work presented herein was performed, for a subset of the authors, as part of their official duties for the United States Government. Funding is hence appropriated by the United States Congress directly.

ACKNOWLEDGMENT

The authors thank D. B. Newell, S. Payagala, G. Fitzpatrick, A. L. Levy, and E. C. Benck for assistance with the internal NIST review process. Commercial equipment, instruments, and materials are identified in this paper in order to specify the experimental procedure adequately. Such identification is not intended to imply recommendation or endorsement by the National Institute of Standards and Technology or the United States government, nor is it intended to imply that the materials or equipment identified are necessarily the best available for the purpose.

The authors declare no competing interests.